\documentclass[a4paper]{article}

\usepackage[margin=1in]{geometry} % full-width

\usepackage{amsmath}
\usepackage{amsthm}
\usepackage{amssymb}

\usepackage[utf8]{inputenc}
\usepackage[hidelinks]{hyperref}
\usepackage{xspace}

\usepackage{placeins}

\newcommand{\Ilim}{I_\text{lim}} % ------------------ Capacitance External
\newcommand{\figref}[1]{Figure \ref{#1}} 
 
\newcommand{\Eqref}[1]{Eq.~(\ref{#1})} 
\newcommand{\Ohm}{$\Omega$\xspace}
\newcommand{\half}{\frac{1}{2}}
\newcommand{\rndP}[1]{\left( #1 \right)}
\newcommand{\Ueff}{U_{\text{eff}}}
\newcommand{\Vth}{V_\text{th}}
\newcommand{\Eth}{E_\text{th}}
\newcommand{\Nef}{N_\text{ef}}
\newcommand{\etaRW}{\overline{\eta}_\text{v}}
\newcommand{\etaV}{\eta_\text{v}}

\newcommand{\Rsf}{R_\text{SF}}
\newcommand{\Rsfmin}{R_\text{SFmin}}
\newcommand{\Rv}{R_\text{v}}
\newcommand{\Rs}{R_S}
\newcommand{\Rc}{R_c} 
\newcommand{\Rx}{R_x} 
\newcommand{\Rt}{R_T}
\newcommand{\RL}{R_L}

\newcommand{\EO}{\mathcal{E}_0}
\newcommand{\Ev}{\mathcal{E}_\text{v}}
\newcommand{\Es}{\mathcal{E}_s}

\newcommand{\Cx}{C_x}
\newcommand{\Ls}{L_\text{S}}

\usepackage{subcaption}
\DeclareCaptionLabelFormat{nocaption}{}
\newcommand{\myfigurewidth}{0.5\linewidth}
\usepackage[sort&compress,numbers,square]{natbib}
\usepackage{graphicx, color}
\graphicspath{{p/}}

\usepackage{mathrsfs} % for \mathscr command

\usepackage{lipsum}

\title{Energy partitioning in electrostatic discharge with variable series load resistor}
\author{\normalsize Claudia A. M. Schrama$^{1,2}$\thanks{Email address: \texttt{cschrama@mines.edu}}, Calvin Bavor$^1$, P. David Flammer$^1$, Charles G. Durfee$^1$}

\date{\normalsize
	$^1$Colorado School of Mines, Physics Department, 1500 Illinois St., Golden, 80401, CO, USA\\%
    $^2$University of Pavia, Department of Physics, via Bassi 6, 27100, Pavia, Italy\\%
    [2ex]%
}

\begin{document}
\maketitle
	
\begin{abstract}
This paper presents an experimental investigation into the energy partitioning of quasi-static electrostatic discharge (ESD) events in air, a scenario in which the discharge occurs across a gap length that can be considered fixed. We systematically characterize the energy transferred to a series victim load across a broad range of resistances (0.1 to 10,000~\Ohm) and circuit parameters, including capacitance and gap length. Our results show that the fraction of stored energy delivered to the victim load is largely independent of gap length. We demonstrate that our extension of the classic Rompe-Weizel spark resistance model effectively predicts the scaling of this energy transfer, establishing a clear link between spark resistance and energy partitioning. These findings provide a predictive framework that should be valuable for guiding safety requirements for energetic materials and ignition scenarios and will inform the development of more accurate circuit models that can be applied to a wider range of ESD events such as those found in the electronics industry.

\vspace{5pt}
\noindent\textbf{Keywords:} Electrostatic Discharge, Energy Partitioning, Rompe-Weizel Model, Breakdown Voltage, Victim Load, Spark Resistance
\end{abstract}

% \tableofcontents

\section{Introduction}

Electrostatic discharge (ESD) is a natural phenomenon that poses a significant threat across numerous high-tech industries, from aerospace to explosives management to semiconductor manufacturing. ESD can rapidly damage or destroy sensitive electronic components through excessive heating from high current densities or dielectric breakdown from intense electric fields \cite{Smallwood2020, Greason1984}. As modern electronic systems become increasingly miniaturized and sensitive, understanding and mitigating these destructive effects is crucial. ESD also poses ignition risks for energetic materials or can degrade them over time, leading to safety hazards and substantial economic losses~\cite{Lovstrand1981, Guoxiang1982, Rizvi1992, Hearn2001, Talawar2006, Larson1989, Ohsawa2011, Weir2013}. 
Ignition may occur through direct contact between an air spark and flammable material, or via current delivered to a secondary victim, causing a chain reaction. In either case, understanding how the initial stored energy is divided between the spark and any series resistive load is important. 

Researchers have extensively investigated various aspects of ESD to unravel its physical mechanisms. For instance, studies on spherical-plane geometries have explored spark dynamics~\cite{Zaridze1996, Bach1981, Hearn2005, Mori2007, Chubb1982}, while ESD from charged human bodies has received specific attention due to its everyday relevance in electronics handling~\cite{Katsivelis2015, Taka2009}. Other investigations have focused on transient effects, such as the influence of approach speed on peak currents and rise times~\cite{Yoshida2012, Daout1987, Tomita2015}, or the impact of rapid voltage rise times on discharge characteristics \cite{Parkevich2019}. These works focus on various aspects of ESD formation, with no specific emphasis on understanding how the energy delivered to resistive circuit elements in series with the ESD changes with circuit parameters.

In recent work by our group~\cite{SCHRAMA2025104205}, we explored ESD energy partitioning in a system where a spark channel was formed between two conducting electrodes in series with a `victim' load resistor. 
The resistance of the victim load in that previous study was specifically chosen to be small relative to the lowest resistance of the spark channel. In that framework, the vast majority of the initial stored energy was delivered to the spark plasma. We investigated how various circuit parameters — capacitance, inductance, electrode geometry and gap length — influenced the fraction of the stored energy that is dissipated by the victim resistance. We showed that the fraction of stored energy delivered to the victim load was simply proportional to the product of the system capacitance and the victim load resistance. This scenario corresponded to an application where an ESD event might deliver current through a low-resistance detonator bridgewire. 

In this paper, we extend our analysis to a wider range of victim load resistances to open the applicability of our model to more diverse components. The application scenario we are addressing is a case where current is delivered directly through an energetic material, in which case the victim load may have much larger resistance than considered previously. We use a range of victim load resistances spanning five orders of magnitude, from 0.1 to 10,000~\Ohm.  This broad parameter space allows for a detailed study of how energy transfer to a victim resistance scales with its impedance. 

Given that the victim load is in series with the spark gap, we expected that, for large victim load resistances, most of the stored energy would be delivered to the victim load.
While we found that this is indeed the case, we also found that the nonlinearity of the spark resistance with respect to the current leads to a coupling of the spark resistance to the victim load.  As in our previous work, we adapt the classical Rompe-Weizel spark resistance model~\cite{Rompe1944}, showing that we can accurately predict this energy transfer across different physical regimes. We note that while we focus on the partitioning of the initial stored energy to the spark and victim loads, we will show that the time scale of that energy delivery also varies with the size of the victim load. We anticipate that these results will find broad applicability to safety scenarios. We also note that the details of how a device or system is sensitive to the ESD current pulse must be incorporated into a full safety assessment. For example, it has been shown that for energetic materials, their sensitivity to ESD damage varies with a number of parameters, including chemical composition, mechanical properties, temperature, and moisture content~\cite{Talawar2006}.

While the risk surrounding energetic materials is the application driving our work, the framework we present here should have broader applicability. For example, a series resistive load may be added to a circuit to reduce energy delivered to either the spark (of interest for ignition/combustion) or to another sensitive component in series with the spark. While the inductances and capacitances relevant to ESD events around electronic components are much smaller than those studied here, we anticipate that the framework we present here, which accounts for the nonlinear spark resistance, will also find application in that area. 

\section{Materials and Methods}
\begin{figure}[ht] 
    \centering
\includegraphics[width = 0.57\linewidth]{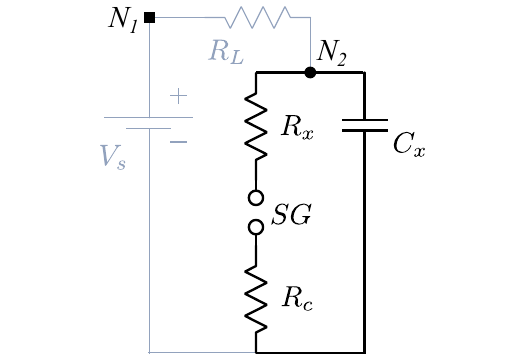}
    \caption{ESD circuit diagram highlighting the charging circuit (thin, light blue) and the discharge circuit (black). The circuit nodes $N_1$ and $N_2$ shows the possible placements of the HVP.}
    \label{fig-circuit-diagram}
\end{figure}

Our experimental setup closely follows the methods presented in our previous work \cite{SCHRAMA2025104205}.
The experimental setup mimics ESD events with slowly closing gaps between conducting objects, ensuring the gap separation is effectively constant during the breakdown process.
\figref{fig-circuit-diagram} illustrates the circuit layout, which is comprised of a charging and a discharge circuit.
The external capacitor, $\Cx$ (TDK low inductance UHV series), placed in parallel with the spark gap branch, is charged by a Spellman SL30PN300 high-voltage power supply through a 100~M\Ohm current-limiting resistor, $\RL$. For data acquisition with the 100~pF $\Cx$, $\RL$ was switched to a 1~G\Ohm resistor to keep a relatively consistent charging time. 

The ESD circuit consists of a spark gap ($SG$) in series with a low-inductance resistor $\Rx$ (Vishay Dale) and current viewing resistor (CVR) $\Rc$ (SSDN- series from T\&M Research Products). As values of $\Rx$ were varied, we selected different values of $\Rc$ to be either 0.0983~\Ohm or 0.5116~\Ohm according to the peak current value. The `victim' load is defined as the sum $\Rv = \Rc + \Rx$.

\begin{figure}[ht]
    \centering
\includegraphics[width = \linewidth]{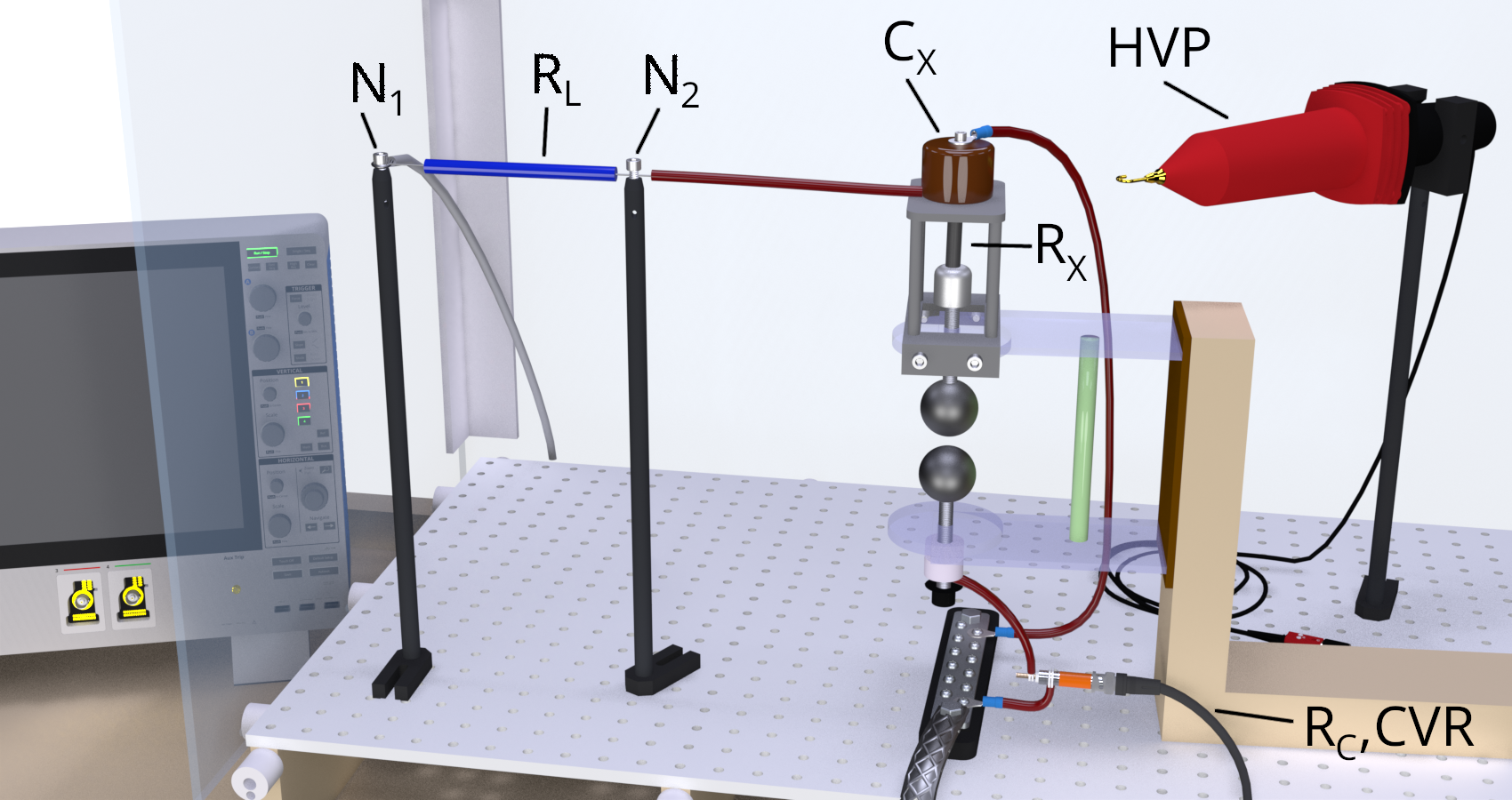}
    \caption{Spatial visualization of the OAS discharge circuit, highlighting the $\Cx$, $\RL$, $\Rx$, $\Rc$, and HVP circuit elements. $N_1$ and $N_2$ are the circuit nodes where the HVP can be connected for voltage monitoring.}
    \label{fig-oas-circuit}
\end{figure}

A Cal Test Electronics CT4028 high-voltage probe (HVP), with a 900~M\Ohm input resistance and 220~MHz bandwidth, was used to measure the voltage of the discharge events. The HVP was positioned at one of two circuit nodes: $N_1$ (pre-$\RL$) or $N_2$ (post-$\RL$), as labeled in \figref{fig-circuit-diagram} and \ref{fig-oas-circuit}.
The HVP's primary function at the $N_1$ location was to accurately capture the breakdown voltage of the discharge. To measure the time-dependent voltage traces — a requirement for calculating the time-dependent spark resistance — the HVP was relocated to the $N_2$ position.

\begin{figure}[ht]
\captionsetup[subfigure]{labelformat=nocaption}
\centering
\includegraphics[width=\myfigurewidth]{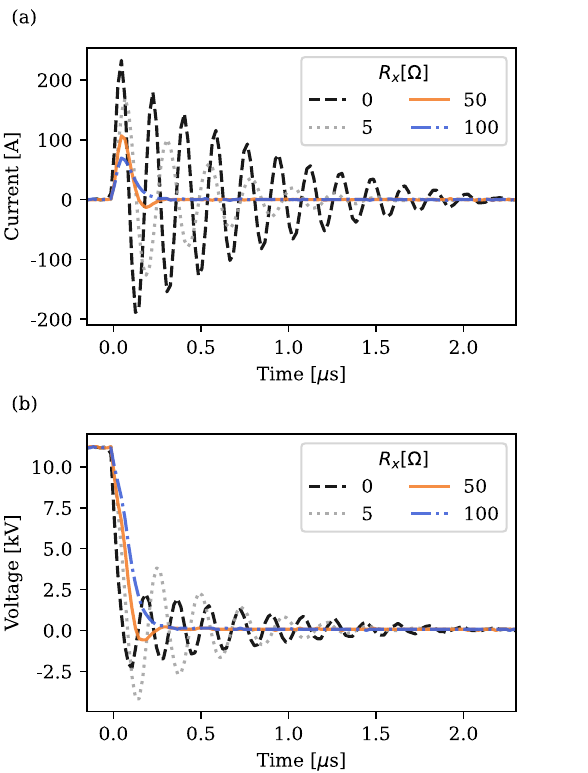}
\begin{subfigure}{0\linewidth}\caption{}\label{fig:3a}\end{subfigure}
\begin{subfigure}{0\linewidth}\caption{}\label{fig:3b}\end{subfigure}
\caption{Voltage and current traces for ESD in the OAS with $\Cx = 700$~pF, $\Rc = 0.0983$~\Ohm, 3.81~mm gap, and varying $\Rx$. The system transitions from underdamped to overdamped behavior as $\Rx$ increases.}
\label{fig:3}
\end{figure}

The circuit was simulated using LTspice, where the spark was modeled as a capacitor in parallel with a time-dependent resistance given by \Eqref{eq:RWresistance}~\cite{Jobava2000}. The simulations revealed that the combined resistance and capacitance of the HVP coupled with a large victim load resistance, $\Rv>100$~\Ohm, significantly interfered with the circuit dynamics when the probe was located at $N_2$. Specifically, this interference caused the current measured by the CVR to differ from the current passing through $\Rx$. This disagreement for large values of $\Rx$ would lead to inaccurate calculations of the circuit's energy dynamics. 
Consequently, the majority of the data presented in this paper was measured with the HVP placed at $N_1$, ensuring that the HVP's internal circuit did not interfere with the overall discharge dynamics. For the specific spark resistance measurements where $\Rx \leq 100$~\Ohm, the HVP's influence was minimal, allowing data collected at $N_2$ to be reliably used for the time-dependent resistance calculation. Data presented in \figref{fig:3}--\ref{fig:5} and \ref{fig:12a} show data where the voltage was collected at $N_2$, all other data was collected with the HVP located at $N_1$.

We constructed an open-air system (OAS) to generate ESD events, see \figref{fig-oas-circuit}.
The OAS spark gap, a modified Ross Engineering SG-40-H commercial high-voltage spark gap, features two 3.75~cm diameter graphite spherical electrodes with a variable gap length. 
We estimated the capacitance of the electrode systems to be below 10~pF. The OAS system's inductance, with $\Rx=0$, was inferred to be $L_\text{OAS}=1.1\pm0.2$~$\mu$H from the discharge current's ring-down frequency. 
This relatively high inductance stems from the large current loop formed by the return wire between $\Cx$ and the ground terminal. 
Finite element simulation, using Comsol Multiphysics, of the discharge circuit yielded an inductance of 0.98~$\mu$H, which aligns well with the measured value.  All data presented were collected in air at standard pressure for Golden, Colorado (approximately 630~torr or 0.83~atm).

Representative voltage and current traces from the HVP and CVR are shown in \figref{fig:3}. The data sets were all taken with a 3.81~mm gap and 700~pF $\Cx$, while varying $\Rx$ from 0 to 100~\Ohm. The current plots show an initial rapid spike, followed by oscillations that gradually decrease over time, eventually returning to 0~A.

The voltage traces show an initial rapid drop from the breakdown voltage (11 kV) to a minimum value, followed by oscillations that gradually dampens over time, ultimately settling to 0~V. An increase in the circuit resistance (varying $\Rx$) reduces the amplitude and duration of the oscillations, transitioning the circuit response from underdamped ($\Rx = 0,\ 5$~\Ohm) to critically or overdamped ($\Rx = 50,\ 100$~\Ohm). 

\subsection{Data Processing}
The initial stored energy, $\EO$, in the system is calculated from the breakdown voltage, $V_b$, and the circuit capacitance, $C$,
\begin{equation}
    \EO=\half C V_b^2.
\end{equation}
The circuit capacitance includes $\Cx$ and the capacitance from the electrodes.
The victim energy, $\Ev$, is calculated from the CVR traces. The total energy dissipated by the victim load is obtained by integrating the power, $P=I_v^2\Rv$,
\begin{equation}
    \Ev=\Rv\int_0^\infty I_v^2~\mathrm{d}t. \label{eqn-EnergyVictim}
\end{equation}
Consistent with our previous work \cite{SCHRAMA2025104205}, charge and energy conservation were verified (where data was available) to validate the measurements and probe calibrations.

\begin{figure}[ht]
\captionsetup[subfigure]{labelformat=nocaption}
\centering
\includegraphics[width=\myfigurewidth]{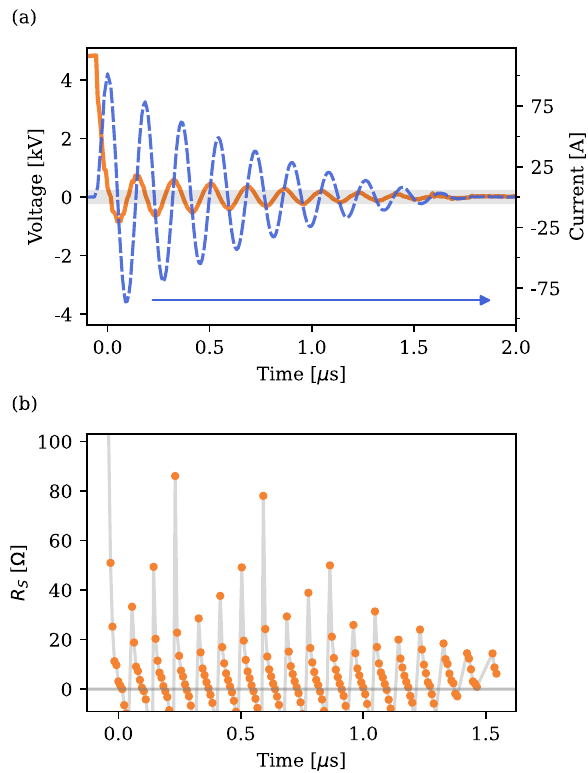}
\begin{subfigure}{0\linewidth}\caption{}\label{fig:4a}\end{subfigure}
\begin{subfigure}{0\linewidth}\caption{}\label{fig:4b}\end{subfigure}
\caption{Unadjusted (a) current and voltage traces for (b) spark resistance measurement. With the OAS, $\Cx = 700$~pF, $\Rv=0.0983$~\Ohm  at 1.27~mm gap. The highlighted region in (a) indicates the $\Ilim$ region and applies to the current trace only.}
\label{fig:4}
\end{figure}

When data is collected with simultaneous time-dependent voltage (HVP at $N_2$ in \figref{fig-circuit-diagram}) and current measurements, we calculate the total time-dependent resistance, $\Rt(t)$. The total resistance includes the spark resistance and the victim load, $\Rt(t) = \Rs(t) + \Rv$. 
To ensure an accurate calculation of $\Rt(t)$, it is essential to account for the cable delay difference between the measurement locations of the HVP and CVR data and the oscilloscope, as detailed in \cite{SCHRAMA2025104205}. 
\figref{fig:4} illustrates the resistance calculation based on Ohm's Law ($R = V/I$) and highlights a significant phase shift between the current and voltage traces that \textit{cannot} be attributed to the measured cable delay. Direct division of the voltage and current traces leads to spikes in the resistance measurement. 

To accurately determine the spark resistance, we must account for the voltage drop across inductive elements, which arises from the inherent circuit geometry and current paths. We subtract this inductive voltage drop from the measured voltage before applying Ohm's Law \cite{Castera2014}:
\begin{equation}
    \Rt(t) = \frac{V(t) - \Ls \frac{d I(t)}{d t} }{I(t)} .\label{eqn-Rs_from_VandI}
\end{equation}
Here, $V(t)$ is the HVP trace, $I(t)$ is the CVR current, and $\Ls$ is the effective inductance. $\Ls$ is a subset of the full circuit inductance, encompassing the inductance of the spark and victim load branch, not making a distinction between which circuit elements cause the shift in the voltage signal. The inductance parameter is determined during data analysis to align the zero-crossing points in the voltage and current traces. The $\Ls$ value is constant for a given circuit layout (same $\Cx$, $\Rc$, $\Rx$ while varying gap lengths) and ranges between 330~nH to 1250~nH, while varying $\Cx$ and $\Rx$.  The range of $\Ls$ due to the different $\Rx$ elements placed in the circuit. 
Our LTspice model provided insight into the observed voltage and current shifts, allowing us to accurately align the data using the $\Ls$ parameter. This is particularly critical for data collected when the system is critically or overdamped, as these conditions eliminate the zero crossings typically used for alignment.

\begin{figure}[ht]
\captionsetup[subfigure]{labelformat=nocaption}
\centering
\includegraphics[width=\myfigurewidth]{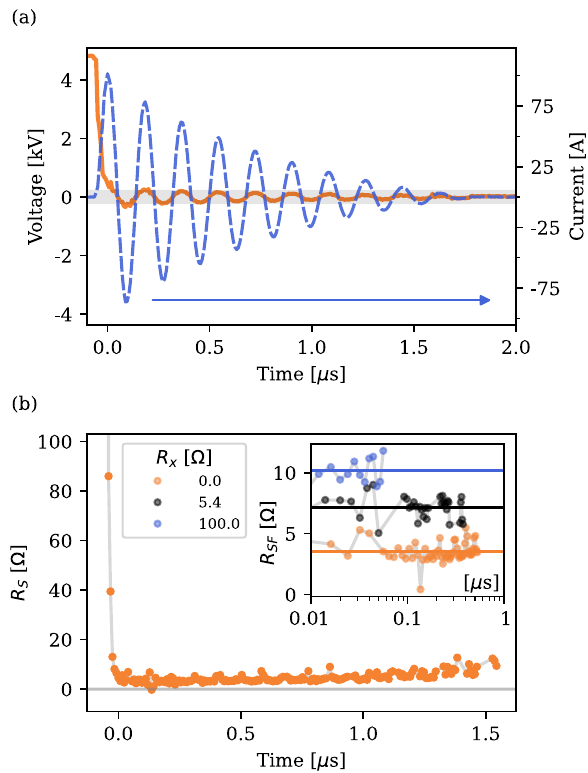}
\begin{subfigure}{0\linewidth}\caption{}\label{fig:5a}\end{subfigure}
\begin{subfigure}{0\linewidth}\caption{}\label{fig:5b}\end{subfigure}
\caption{Adjusted (a) current and voltage traces and (b) measured spark resistance. With the OAS, $\Cx = 700$~pF, $\Rc=0.0983$~\Ohm at 1.27~mm gap. The highlighted region in (a) indicates the $\Ilim$ region and applies to the current trace only.}
\label{fig:5}
\end{figure}

To minimize noise in the resistance calculation, data is only considered when the absolute current, $\left|I(t)\right|$, exceeds a predetermined limit, $\Ilim$, as indicated by the shaded regions in \figref{fig:4a} and \ref{fig:5a}. 

\figref{fig:5a} shows the calculated spark resistance using \Eqref{eqn-Rs_from_VandI}. The spark resistance starts by dropping quickly at the onset of the discharge and then remains nearly constant for the remainder of the measurable discharge current. This resistance is calculated for the specific experimental conditions with $\Rv =0.0983$~\Ohm, $\Cx=700$~pF and a 1.27~mm gap in the OAS. The inset plot in \figref{fig:5b} shows the method of determining the final spark resistance $\Rsf$, which is an average over multiple data points that are collected right after the initial voltage drop. 

\FloatBarrier
\section{Theory \label{sec-theory}}

To interpret the measured energy transfer to the varying victim load, we build upon our previously presented theoretical framework \cite{SCHRAMA2025104205}. This framework primarily utilizes the nonlinear spark resistance model developed by Rompe and Weizel \cite{Rompe1944}. 
The basic assumption of the Rompe-Weizel (RW) model is a constant ``effective'' energy cost per electron created during the breakdown process, $\Ueff$, and constant electron mobility, $\mu$. This, combined with an assumption that the threshold voltage is proportional to the gap length, $h$, and electric field, $\Vth = h\Eth$, leads to a simple model for the energy delivered to the circuit components; the electric field strength for breakdown, $\Eth$, can be determined from Paschen curves \cite{Paschen1889,Raizer1991}.

The actual energy required to generate electrons in a low temperature plasma, $\Ueff$, significantly exceeds the ionization energy of the molecules in the background gas. For this study, we treat $\Ueff$ as an empirical parameter that accounts for various energy channels, including ionization, dissociation, particle heating and excitation.

As in our previous work \cite{SCHRAMA2025104205}, we model the plasma channel as a uniform cylinder with length $h$, cross-sectional area $A$ and resistance $\Rs = h/(\sigma A)$. The conductivity is $\sigma = n_e e \mu$, where $n_e$ is the electron number density. The energy expended to create electrons comes from the Joule heating, $\vec{J}\cdot \vec{E}$,  
\begin{equation}
    \Ueff\frac{d n_{e}}{d t} = \vec{J} \cdot \vec{E} = \frac{J^{2}}{\sigma}=\frac{J^{2}}{e\,\mu\,n_e},
\end{equation}
which represents the energy transfer from the electromagnetic field to the plasma.
This equation can be integrated to determine the time dependence of the electron density, $n_{e}$, which then goes into the expression for the spark resistance. Finally, writing the current density as $J=I/A$, where $I$ is the current delivered through the spark, we arrive at an expression for the nonlinear spark resistance:
\begin{equation}
    \Rs(t) = 
    \rndP{
    \frac{2\, a_R}{h^2} \int_0^t I(t')^2 \mathrm{d} t
    }^{-1/2}.
\label{eq:RWresistance}
\end{equation}
Here, we define the Rompe-Weizel constant, $a_R \equiv e \mu/\Ueff$. Note that the spark resistance in this model is independent of the cross-sectional area, $A$, and it only depends on the total number of electrons produced. Since $\Ueff$ and $\mu$ are challenging to measure independently, $a_R$ is treated as the empirical parameter of interest. 
Values for the Rompe-Weizel constant, $a_R$, typically range from 0.5 to 2~cm$^2/$sV$^2$ in air~\cite{Jobava2000}, a range that is consistent with the findings reported in our previous investigation~\cite{SCHRAMA2025104205}. 

\subsection{Energy partitioning between spark and victim load}
In this research, our primary interest lies in the energy delivered to a resistive victim load in series with the spark. Assuming a uniform electric field, the initial stored energy increases quadratically with $h$:
\begin{equation}
    \EO= \half C \Vth^2 = \half C \Eth^2 h^2.
    \label{eq:idealE0}
\end{equation}

Within the RW model, the energy deposited into the spark, $\Es$, is defined as $\Es = \Nef\Ueff$, where $\Nef$ is the total number of electrons produced in the spark.
Therefore, the final spark resistance, $\Rsf$, after the arc, is a function of the energy delivered to the spark,
\begin{equation}
    \Rsf = \frac{h^2}{\Nef\, e\, \mu} = \frac{h^2}{\Es\, a_R}. \label{eqn-Rsf}
\end{equation}
For convenience, we define a minimum spark resistance, 
\begin{equation}
    \Rsfmin = \frac{2}{a_R\, C\, \Eth^2}, \label{eqn-Rsfmin}
\end{equation}
occurring when all of the initial stored energy is consumed in producing electrons $(\Es = \EO)$.
In this limit, the final resistance is independent of the spark gap separation, $h$.
 
The energy invested in forming the spark, $\Es$, can be calculated through similar reasoning for \eqref{eq:RWresistance}, where $\Nef$ is the integral of $n_e(t)$ times the spark volume $h\, A$, leading to the following differential equation solution:  

\begin{equation}
    \Es = \sqrt{\frac{2 h^2}{a_R} \int_0^{t_f} I^2(t) dt}.
    \label{eqn-EnergySpark}
\end{equation}

The energy dissipated by the victim load, $\Ev$, is determined by integrating the power, as presented in \Eqref{eqn-EnergyVictim}.  By substituting $\Rsfmin$ from \Eqref{eqn-Rsfmin} and $\Ev/\Rv$ from \Eqref{eqn-EnergyVictim} into \Eqref{eqn-EnergySpark}, the spark energy can be expressed as $\Es = \sqrt{2\Rsfmin\EO\ {\Ev/\Rv}}$. Applying energy conservation ($\EO=\Es+\Ev$), we arrive at
\begin{equation}
    (1-\etaV)^2 = 2 \etaV \frac{\Rsfmin}{\Rv}. 
\end{equation}
Here, $\etaV = \Ev/\EO$ represents the fraction of initial stored energy dissipated through the victim load. We can solve this quadratic equation, selecting the physically relevant negative root solution that ensures $0\leq \etaV \leq 1$: 
\begin{equation}
    \etaV = 1 + \frac{\Rsfmin}{\Rv} - \sqrt{2 \frac{\Rsfmin}{\Rv} + \rndP{\frac{\Rsfmin}{\Rv}}^2}. 
    \label{eqn-EtaV_in_Rv_Rsfmin}
\end{equation}

\begin{figure}[ht]
\captionsetup[subfigure]{labelformat=nocaption}
\centering
\includegraphics[width=\myfigurewidth]{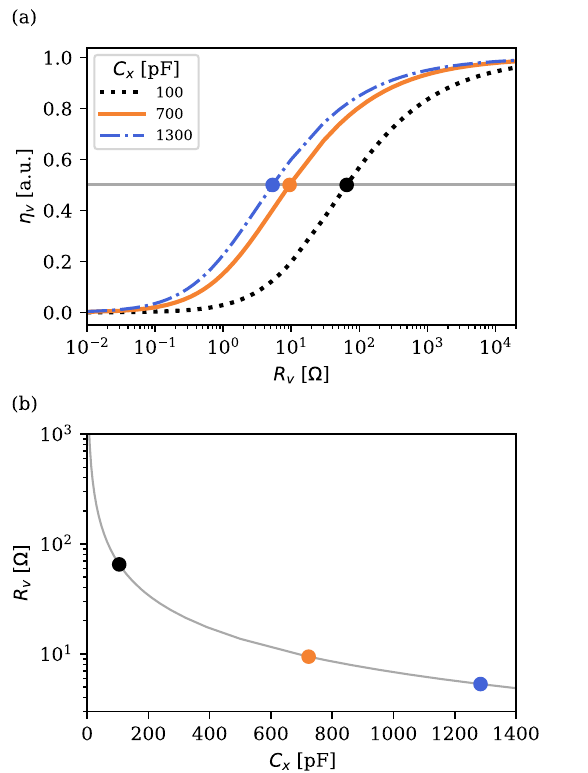}
\begin{subfigure}{0\linewidth}\caption{}\label{fig:6a}\end{subfigure}
\begin{subfigure}{0\linewidth}\caption{}\label{fig:6b}\end{subfigure}
\caption{(a) The fraction of energy dissipated by the victim load from \Eqref{eqn-EtaV_in_Rv_Rsfmin} is shown for three different circuit capacitances, while setting $h=1.27$~mm, $V_b=4.57$~kV, and $a_R=0.5$~cm$^2/$sV$^2$. The value of $\Rv$ along the gray line in (a) is shown in (b) for capacitance values ranging from 1 to 1400~pF.}
\label{fig:6}
\end{figure}

\figref{fig:6a} illustrates the variation of  $\etaV$ across a wide range of victim load resistances for selected external capacitor sizes. The curves use the breakdown voltage and $a_R$ values specified in the caption (a constant $a_R$ implies a fixed ratio of $\mu$ and $\Ueff$). The fraction of stored energy delivered to the victim load is initially small for small $\Rv$, but increases towards 100~\% at larger $\Rv$. The crossover value of $\Rv$ at which $\etaV = 0.5$ is dependent on the external capacitance $\Cx$, which serves as a proxy for the initial stored energy. For larger $\Cx$ in the system, the $\etaV$ balance point is reached at a lower $\Rv$. 
\figref{fig:6b} shows how the $\Rv$ value at this crossover point changes with $\Cx$.

In the limit of small victim load ($\Rv << \Rsfmin$), we can expand \Eqref{eqn-EtaV_in_Rv_Rsfmin} to second order to obtain the approximate form, 
\begin{equation}
    \etaV \approx \frac{1}{2} \frac{\Rv}{\Rsfmin}\left(1-\frac{\Rv}{\Rsfmin}\right).
    \label{eqn:approx-etaV}
\end{equation}
The expression used in our previous work \cite{SCHRAMA2025104205} was the first-order portion of this approximation, i.e. $\etaV \approx \Rv/(2 \Rsfmin)$.
That earlier paper investigated the scaling of $\etaV$ with circuit parameters $\Rv$ and $\Cx$. We demonstrated that, for small victim loads, the energy delivered to the victim scales linearly with both the product of circuit capacitance and victim load resistance:
\begin{equation}
    \etaRW = \frac{\etaV}{C \Rv} = \frac{a_R \Eth^2}{4}.\label{eqn-etaRW}
\end{equation}
This scaling relationship was specifically derived under the assumption of small victim load ($\Rv < 1$~\Ohm). \Eqref{eqn-etaRW} thus provides a direct method for predicting the maximum energy transfer to a series resistance in such low-resistance scenarios. 
We also observed that $\etaV$, and thus also $\etaRW$, remained largely independent of gap lengths exceeding 1~mm.

\begin{figure}[ht]
    \centering
    \includegraphics[width = \myfigurewidth]{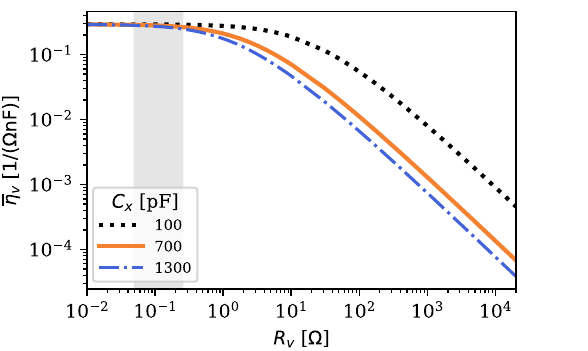}
    \caption{The fraction of energy dissipated by the victim load scaled to the circuit capacitance and victim load resistance from \Eqref{eqn-etaRW} is shown for three different circuit capacitances, while setting $h=1.27$~mm, $V_b=4.57$~kV, and $a_R=0.5$~cm$^2/$sV$^2$. The gray vertical block indicates the resistance ranges we used in \cite{SCHRAMA2025104205}.  }
    \label{fig:7}
\end{figure}

Here, Eq.~\eqref{eqn-EtaV_in_Rv_Rsfmin} is an extension of the previous results to arbitrary values of $\Rv$. \figref{fig:7} illustrates $\etaRW$ for various victim loads, assuming a constant $a_R$. The trend indicates that $\etaRW$ deviates from a constant value when the victim load exceeds 1~\Ohm. Referring to our approximate form for $\etaV$, \Eqref{eqn:approx-etaV}, we can see that when this deviation occurs, it is not safe to assume that $\Rv <<\Rsfmin$. 
In the opposite limit of a large victim load, we find $\etaV \approx 1-\sqrt{2 \Rsfmin/\Rv}$.  At very large $\Rv$, nearly all the initial energy is expected to be absorbed by the victim load, resulting in minimal energy delivered to the discharge event itself.

The Rompe-Weizel constant, $a_R$, is treated as an empirical fitting parameter, used during post-processing to fit \Eqref{eqn-EtaV_in_Rv_Rsfmin} with the experimental data. The derived $a_R$ values contribute to a refined understanding of the discharge process and can enhance the accuracy of our simplified ESD models.

\subsection{Spark Resistance}

\begin{figure}[ht]
    \centering
\includegraphics[width = \myfigurewidth]{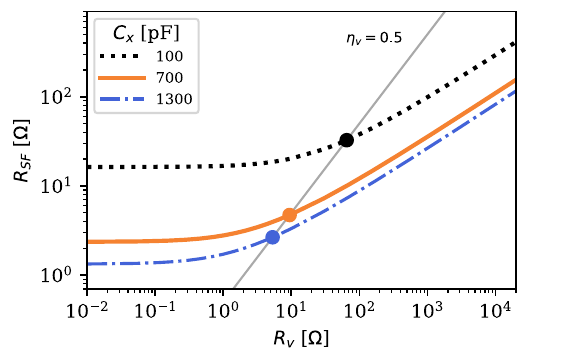}
    \caption{$\Rsf$ from \Eqref{RQ-eqn-RsVsRv} is shown for different $\Cx$ values while setting $\Eth = V_b/h$ with $h=1.27$~mm, $V_b=4.57$~kV, and $a_R=0.5$~cm$^2/$sV$^2$. In these plots, $\Rsfmin$ changes from 17.1 to 2.4 and 1.3~\Ohm from small to large $\Cx$.}
    \label{fig-theory-Rsf}
\end{figure}

The power transfer between the spark and the victim load is dependent on the final spark resistance, $\Rsf$. By combining \Eqref{eqn-Rsf} with preceding equations, we can express $\Rsf$ as
\begin{align}
    \Rsf = \frac{1}{a_R \Cx \Eth^2} \rndP{1 \pm \sqrt{1 +  {a_R \Cx \Eth^2} \Rv}} \label{RQ-eqn-RsVsRv}.
\end{align}
We select the positive root to ensure that when no victim load is present ($\Rv=0$), the final spark resistance equals the minimum possible resistance, $\Rsf=\Rsfmin$. \figref{fig-theory-Rsf} illustrates the trend in the final spark resistance for three different $\Cx$ values. As the victim load increases, $\Rsf$ also increases because the victim load absorbs a larger proportion of the system's energy. Larger values of system capacitance lead to smaller $\Rsf$ and $\Rsfmin$ values at small $\Rv$. Rose et al. \cite{Rose2020} reached a similar conclusion using a non-dimensionalized analysis of the discharge equation.

To calculate $\Rsf$ from experimental data, we combined equations \eqref{eqn-EnergyVictim}, \eqref{eqn-Rsf} and \eqref{eqn-EnergySpark} to express the energy dissipated by the spark as
\begin{equation}
    \Es^2 = 2 \Rsf\ \Es \frac{\Ev}{\Rv}.
\end{equation}
This simplification yields the following expression for $\Rsf$:
\begin{equation}
    \Rsf = \half \Rv \frac{\Es}{\Ev} 
         = \half \Rv \frac{\EO-\Ev}{\Ev}
         = \half \Rv \rndP{\frac{1}{\etaV}-1}. \label{eqn-Rsf-from-RWmodel}
\end{equation}
\Eqref{eqn-Rsf-from-RWmodel} directly provides the spark resistance after initial breakdown, relating it to $\Rv$ and the fraction of energy delivered to $\Rv$. The functional form of $\Rsf$  being independent of gap length suggests that higher stored energy at longer gap lengths is distributed over a proportionally larger spark volume, implying a larger discharge channel radius. 

While traditional voltage dividers suggest $\Rsf = \Rv$ for a 50\% energy dissipation by both resistors, our analysis at the crossover point, where $\etaV = 0.5$, reveals
\begin{equation}
    \Rsf = \half \Rv. \label{eqn-Rsf-fifty-fifty}
\end{equation}
The smallest $\Rsf$ reached is less than $\Rv$. That phenomenon is attributed to the time-dependent nature of the nonlinear spark resistance; specifically, during the initial, bulk energy transfer phase of the ESD, $\Rs(t)$ is larger than $\Rv$. 

\FloatBarrier
\section{Results and Discussion \label{sec-RandD}}

\begin{figure}[ht]
    \centering
\includegraphics[width= \myfigurewidth]{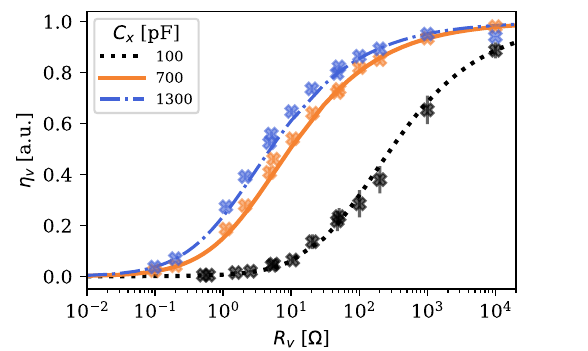}
    \caption{Example $\etaV$ versus $\Rv$ data along with the fitted \Eqref{eqn-EtaV_in_Rv_Rsfmin} for discharges at 3.81~mm gap.  The $a_R$ fit parameter was 0.3789, 1.671, and 1.826~cm$^2/$sV$^2$ for the data taken with $\Cx = $~100, 700 and 1300~pF, respectively.}
    \label{fig:9}
\end{figure}

Data was collected for three different external capacitors (100, 700, and 1300~pF), five different gap lengths (1.27, 2.54, 3.81, 5.08, 6.35~mm), and $\Rv$ values ranging from 0.1~\Ohm to 10~k\Ohm. 
The fraction of energy dissipated by the victim load for ESD events with three different external capacitors at a 3.81~mm gap length are shown in \figref{fig:9}. The crosses indicate the data, and the corresponding curves are fit by varying the $a_R$ parameter.

\begin{figure}[ht]
    \centering
    \includegraphics[width= \myfigurewidth]{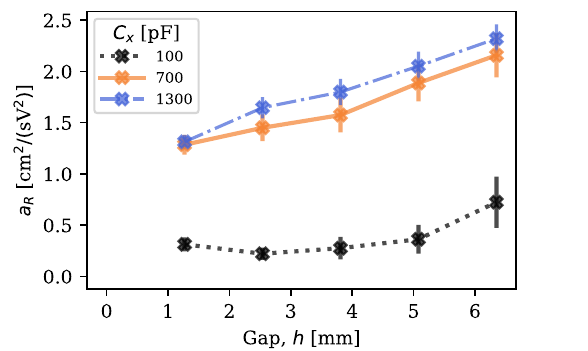}
    \caption{Fit parameter $a_R$, from \Eqref{eqn-EtaV_in_Rv_Rsfmin}, for data taken at various gap lengths and circuit capacitor values. The fit plots in \figref{fig:9}, \ref{fig:11} and \ref{fig:12a} use the average value of these numbers per capacitor value.}
    \label{fig-aR}
\end{figure}

The fit parameter of the model, $a_R$, is plotted in \figref{fig-aR}. For the two larger capacitance values, the data shows that the fit parameter increases with gap length. For the smallest capacitance, $100$~pF, we observe a general upward trend, with a slight initial depression as the gap length increases. The physical meaning of this parameter is being explored further and will be reported on in our future work. For now, we note that these values fall within our own previously measured $a_R$ parameters, and within the range reported by  Jobava et. al. \cite{Jobava2000}.

\begin{figure}[ht]
    \centering
    \includegraphics[width= \myfigurewidth]{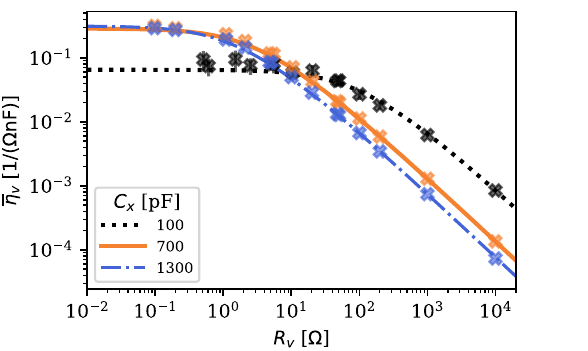}
    \caption{Normalized fraction of energy, $\etaRW$, dissipated by victim load for various resistance with fitted \Eqref{eqn-etaRW}.  The $a_R$ fit parameter was 0.3789, 1.671, and 1.826~cm$^2/$sV$^2$ for the data taken at $\Cx =$ 100, 700 and 1300~pF, respectively. The small error bars show the spread of $\etaRW$ across all tested gap lengths.}
    \label{fig:11}
\end{figure}

The normalized energy ratio for the varying victim load is shown in \figref{fig:11}. For each resistance value, data were collected at five different gap lengths. The minimal spread observed in these data points (as indicated by the min-max bars on the plot) confirms that the normalized energy ratio, $\etaRW$, is gap length invariant. As the victim resistance increases in magnitude, the value of $\etaRW$ concurrently decreases.

\begin{figure}[ht]
\captionsetup[subfigure]{labelformat=nocaption}
\centering
\includegraphics[width=\myfigurewidth]{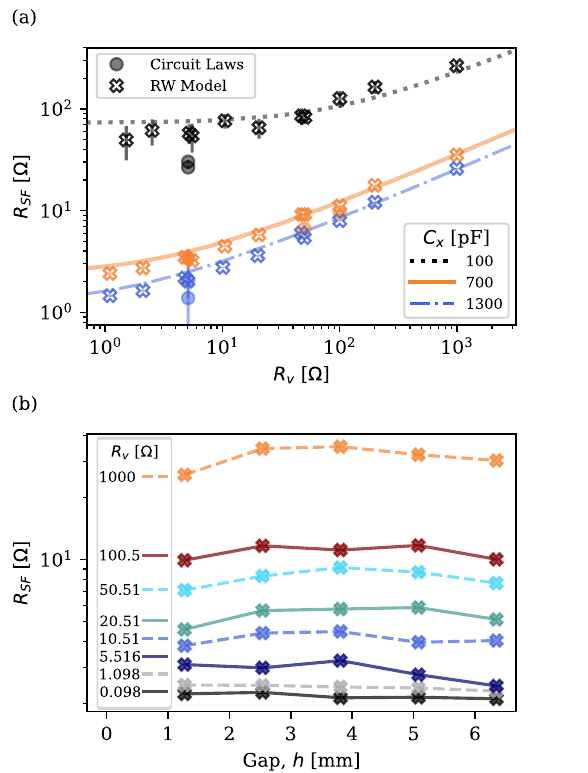}
\begin{subfigure}{0\linewidth}\caption{}\label{fig:12a}\end{subfigure}
\begin{subfigure}{0\linewidth}\caption{}\label{fig:12b}\end{subfigure}
\caption{The final spark resistance $\Rsf$ (a) calculated as a function of victim load resistance for various $\Rv$--$\Cx$ combinations overlayed on top of their corresponding \Eqref{RQ-eqn-RsVsRv} fit. The $a_R$ fit parameter was 0.3789, 1.671, and 1.826~cm$^2/$sV$^2$ for the data taken at $\Cx =$~100, 700 and 1300~pF, respectively. The crosses represent the $\Rsf$ calculation using \Eqref{eqn-Rsf-from-RWmodel}. There are a select number of points, the circles on the plot, in which \Eqref{eqn-Rs_from_VandI} is also used to calculate the resistance. (b) $\Rsf$ as a function of gap length at select $\Rv$ values for $\Cx =700$~pF. There is neither a clear upward nor a downward trend in the $\Rsf$ over gap length, indicating it is gap length independent.}
\label{fig:12}
\end{figure}

We calculated the resistance of the spark in two different ways: time-dependent measurements and the RW model approximation. The resulting resistance values are shown in \figref{fig:12a}, where the crosses represent the $\Rs$ calculated from the RW model approximation, \Eqref{eqn-Rsf-from-RWmodel}, and the circles are calculated using the voltage and current traces, \Eqref{eqn-Rs_from_VandI}. The data is compared to the model version of the resistance from \Eqref{RQ-eqn-RsVsRv}. We see that both measurement types agree with the model, which assumes a constant $a_R$. 

\figref{fig:12b} presents the spark resistance measurements across various gap lengths and different $\Rx$ values, obtained using a 700~pF $\Cx$. The data consistently show that the spark resistance is independent of gap length. The simple resistance formula from Eq.~\eqref{eqn-Rsf} dictates that resistance should have gap length dependence. The independence, therefore, suggests a compensation mechanism within the discharge channel, indicating that the energy is distributed over a larger effective volume to balance the increasing stored energy. In future work, we will further investigate this through direct optical imaging to verify if the channel radius is truly changing.

Our results show that the modified RW model surprisingly captures energy transfer well (as evidenced by the observed behaviour of $\eta$ and $\etaRW$) across a wide range of capacitances and series resistances. While the ranges studied here will be directly applicable to a number of applications, it is important to note that in the microelectronics industry, the charged device model (CDM) \cite{ANSI2025Report}, 
the inductances are typically in the nH range and capacitances range from 7 to 55~pF, leading to much shorter current risetimes (400~ps) than seen in our experiments. In our earlier work \cite{SCHRAMA2025104205}, 
we directly compared ESD events on the system used here to a separate system with an inductance that was an order of magnitude smaller (120~nH) and the fraction of energy delivered to the victim load was not appreciably affected. We also note that our experiments span regimes from over-damped to critically-damped and under-damped, supporting the conclusion that the results are insensitive to inductive effects.

\begin{figure}[ht]
    \centering
    \includegraphics[width=\myfigurewidth]{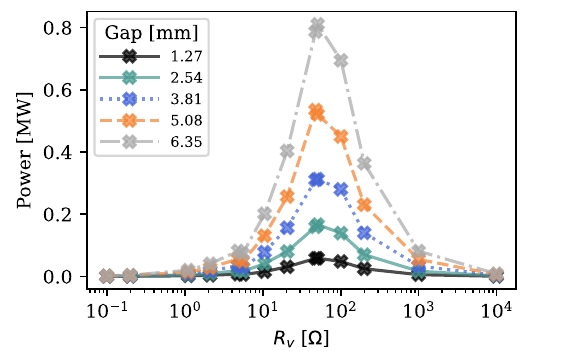}
    \caption{Average power dissipated by the victim load with $\Cx = 700$~pF, for varying victim loads at various gap lengths. The average power was calculated by dividing the energy dissipated by the victim load by the effective duration of the current pulse $\tau_p$, see Eq.~\eqref{eq-currentpulseduration}.}
    \label{fig-PowerTransfer}
\end{figure}

Users implementing this information for safety-critical applications must determine whether their system is energy- or power-limited, as the discharge dynamics govern the power transfer. The effective duration of the power delivery is relatively longer when the pulse is underdamped (small $\Rv$) or overdamped (large $\Rv$), and is shortest for moderate victim loads ($10 < \Rv < 200$~\Ohm) where the current pulse is close to critically-damped (see \figref{fig:3a}). To explore this, in \figref{fig-PowerTransfer} we plot the variation of average power with $\Rx$ and gap length. To obtain the average power, we calculate the energy dissipated by the victim load, then divide by the effective duration of the current pulse, which in turn is calculated by taking the second moment of the current pulse with respect to time: 
\begin{equation}
    \tau_p = \sqrt{\frac{\int (t-T_C)^2 I^2(t) \text{ d}t}{\int I^2(t) \text{ d}t}}. 
    \label{eq-currentpulseduration}
\end{equation}
Here $T_C$ is the time-centroid of the power $I(t)^2 \Rv$. We further normalize the second moment to mimic a power dissipation given by a square current pulse. \figref{fig-PowerTransfer} shows that the power transfer to the victim load peaks for load resistances in the 20-200~\Ohm range. This type of analysis could be important for victim loads that are coupled to dissipative loss mechanisms such as thermal conduction. 

\FloatBarrier
\section{Conclusions}
This study provides a comprehensive investigation into the energy partitioning between the spark channel and a variable series resistive load with quasi-static ESD events. By extending our previous analysis to victim load resistances spanning five orders of magnitude (0.1~\Ohm to 10~k\Ohm), we have provided a detailed, predictive framework for energy transfer scaling with its dependence on key circuit parameters.

Our findings confirm that the fraction of stored energy delivered to the victim load, $\etaV$, is directly tied to the victim load resistance and initial stored energy.
We demonstrate that the simple, nonlinear spark resistance model proposed by Rompe and Weizel can be extended to predict the scaling of this energy transfer beyond the small victim load resistance limit. The model shows that for small victim loads, the energy transfer scales linearly with the product of capacitance and resistance, providing a framework for engineers designing for low-resistance scenarios. For larger victim loads, our data and model reveal a transition in which nearly all the initial energy is absorbed by the load, with the spark resistance increasing as the victim load increases. Fitting the modified RW model to the data, we find $a_R$ values ranging from 0.22 to 2.32~cm$^2/$sV$^2$, which aligns with previously observed values~\cite{SCHRAMA2025104205,Jobava2000}.

While we have been treating the total passive resistance that is in series with the spark as the ``victim'' load, our approach here can apply to scenarios where a secondary resistance is placed in series to dissipate stored energy, diverting energy from the spark (ignition scenarios) or from a smaller-resistance sensitive victim load. Indeed, this is an established control used in the microelectronics industry for circuit protection~\cite{Jack2021,ANSI2025Report}). The results of this paper show that, from an energy-delivery standpoint, one wants to minimize the resistance the victim makes up of the total resistance involved in the ESD event.

This research fills a critical gap in the understanding of ESD by explicitly focusing on the effects on a series resistive element. 
By quantifying the energy transferred to a victim load, our work provides a valuable tool for guiding safety requirements for sensitive electronic components and energetic materials. 
The insights gained from this study, particularly the relationship between victim resistance and energy partitioning, can be used to mitigate the destructive effects of ESD. Furthermore, our findings will help guide the development of future models aimed at predicting circuit behavior under these ESD conditions, thereby enhancing the safety and reliability of modern electronic systems.

\section{Acknowledgments}
We gratefully acknowledge financial support for this work from Los Alamos National Laboratory. We also acknowledge very useful conversations with Jonathan Mace, Dan Borovina, Travis Peery, John Rose, and Francis Martinez at Los Alamos National Laboratory. 

\bibliographystyle{ieeetr} 
\bibliography{biblio}

\end{document}